\documentclass[11pt,a4paper]{article}
\usepackage{jcappub}

\usepackage{amsmath,epsfig}
\usepackage{amssymb,amsfonts}
\usepackage{latexsym}

\relax

\usepackage{graphicx}
\usepackage{color}

\newcommand{\ba}{\begin{eqnarray}}
\newcommand{\ea}{\end{eqnarray}}
\newcommand{\be}{\begin{equation}}
\newcommand{\ee}{\end{equation}}

\newcommand{\pd}{\partial}

\newcommand{\diag}{\mathop{\mathrm{diag}}\nolimits}

\newcommand{\Cc}{\mathcal{C}}
\newcommand{\Dc}{\mathcal{D}}
\newcommand{\Fc}{\mathcal{F}}

\newcommand{\Kc}{\mathcal{K}}
\newcommand{\Lc}{\mathcal{L}}

\newcommand{\const}{\text{const}}

\title{Stable analytic bounce in non-local Einstein-Gauss-Bonnet
cosmology}
\author{Alexey S. Koshelev}
\affiliation{Theoretische Natuurkunde, Vrije Universiteit Brussel and The
International Solvay Institutes, Pleinlaan 2, B-1050 Brussels, Belgium}
\emailAdd{alexey.koshelev@vub.ac.be}

\abstract{
We consider the most general quadratic in curvature stringy motivated non-local
action for the modified Einstein's gravity. We present exact analytic
cosmological solutions including the bouncing ones and develop the relevant
techniques for the further study of this type of models. We also elaborate on
the perturbation formalism and argue that the found bouncing solution is stable
during the bounce phase.
}

\begin{document}
\maketitle
\section{Introduction}

Recent observations~\cite{WMAP} strongly support that primordial inflation is
the theoretical explanation of how the currently observed Universe was formed
at the early stages. Alongside with observations theoretical approaches also
show how nice inflation can be connected to the nucleosynthesis and subsequent
appearance of the particle standard model. A number of inflationary scenarios
are reviewed in~\cite{Mazumdar:2010sa} and
references therein, for instance.

Even though inflation is a great model for many reasons it has problems one of
which is the lack of the UV completion. To be more precise it is not UV-complete
in
the framework of the Einstein's General Relativity (GR) since
geodesics are not past-complete. This is a general statement and it is known as
the Big Bang singularity  elaborated
in \cite{Borde,Linde,Guth}. One
can find that alternatives to Big Bang such as ``emergent'' Universe or
bouncing Universe~\cite{emerge} hit the singularity theorem by Hawking and
Penrose~\cite{Hawking} as long as we are in GR and the space-time is of the
Friedmann--Lema\^{i}tre--Robertson--Walker (FLRW) type.

One of the possible resolution is a modification of GR. This can be done in
general in a number of ways and one may have an insight in this using the review
paper~\cite{Koivisto-rev} and
references therein, for example. It is inevitable that any modification
of gravity introduces higher derivatives and only special structures like
Gauss-Bonnet term or Lovelock terms in general~\cite{Koivisto-rev,lovelock}
preserve the second order of the equations of motion but this is applicable only
in more than $4$ dimensions. On the other hand finite higher derivatives lead
to ghosts due to the Ostrogradski theorem~\cite{ostrogradski}. Having all
orders of higher derivatives may open a way to evade the Ostrogradski
statement and a successful attempt in this direction was
made considering a special class of gravity modifications where
higher curvature corrections are accompanied with non-local
operators~\cite{BMS,BKM}. Analysis in those papers shows how one can
construct a ghost-free and asymptotically free modification of GR featuring
a non-singular bouncing solution, and which resembles the GR itself in the IR
limit.\footnote{Similar approaches involving non-local models were used in
other
cosmological and gravity contexts in the
literature~\cite{inflation,sgc,nlgravity,ArkaniHamed:2002fu,Barvinsky2003}.}.
The
further study of this model~\cite{last} has shown that the model
features the
expected perturbative spectrum at late times and is stable with respect to small
isotropic inhomogeneous perturbations during the bounce phase and in parallel a
more general model~\cite{BGKM} (see also \cite{newBKM}) was proposed and
considered in the Minkowski
background.

Absolutely the non-local operators is what makes these models novel and the
operators in question are of type of analytic functions of the covariant
d'Alembertian operator, i.e. $\Fc(\Box)$.\footnote{Theoretically motivated
operators such as $1/\Box$ were considered, for instance,
in~\cite{One_over_Box,oobBI} and references therein.}
We find it is quite important that the initial inspiration for introducing these
operators came from string field theory (SFT) models which as the whole theory
is a UV-complete
non-perturbative description of strings. We refer the reader to more stringy
oriented literature~\cite{sft_review,padic_st,ZS,douglas,marc,sft} for a more
comprehensive overview of this aspect. A decent progress was achieved in
studying non-local scalar field models derived from SFT in the cosmological
context~\cite{Non-local_scalar}-\cite{KV} as well as exploration of other
aspects of this type of models including their
thermodynamics~\cite{linearized,BJK,abe}.
The major question of rigorous derivation of the modified GR action involving
the non-local operators of interest from the scratch, i.e. from the closed SFT,
is still awaiting for its resolution but
this is beyond the scope of our present study. 

The present paper is aimed at extending and continuing both
papers~\cite{last,BGKM}. In~\cite{last} a lot of technical issues were solved
for a model which contains the scalar curvature squared non-local term. The
main focus there is perturbations around
a bouncing solution. In~\cite{BGKM} non-local terms containing the Ricci and
Riemann
tensors squared were added but only analyzed around the Minkowski background. We
literally want to join those works in the present paper. We confine ourselves by
considering the FLRW type of the metric and positive cosmological term
$\Lambda$. Having these in mind we focus on deriving full equations of motion
and bringing them to a form one can use in the future study. Our main goal is to
present here classical solutions and develop the perturbation technique at least
in some regimes or around some backgrounds. This is crucial for claiming the
model is viable or not for the purpose of describing the non-singular bounce. 

The paper is organized as follows: In Section~2, we formulate the model
and introduce the relevant notations. Also we write down full equations of
motion relevant for the FLRW type metric. In Section~3, we demonstrate
explicitly that one of the known bouncing solutions, namely the cosine
hyperbolic, is the solution to our extended equations of motion as well. In
Section~4, we perform an attempt to construct more solutions and develop
relevant
techniques. In
Section~5, we are building the road to analyzing perturbation. We find out
we are lucky to draw the stability statement about our exact bouncing solution
from general arguments as well as sketch the derivation of the closed
system of perturbation
equations for the de Sitter asymptotic. In Section~6, we summarize what is done
and formulate
open questions for the future study.

\section{Action and equations of motion}

We focus on the model described by the following non-local action
\begin{equation}
S=\int d^4x\sqrt{-g}\left(\frac{M_P^2}2R+\frac{\lambda}2\left(R\Fc_1(\Box) R+R^\mu_\nu\Fc_2(\Box)R^\nu_\mu+C_{\mu\nu\alpha\beta}\Fc_4(\Box)C^{\mu\nu\alpha\beta}\right)-\Lambda\right)
\label{model}
\end{equation}
where we limit ourselves with $O(R^2)$ corrections.
Here the dimensionality is manifest and in the sequel all the formulae are
written having $4$ dimensions in mind, $M_P$ is the Planckian mass, $\Lambda$ is
a cosmological constant and $\lambda$ is a dimensionless parameter measuring the
effect of the $O(R^2)$ corrections. The most novel and crucial for our analysis
ingredients are the functions of the covariant d'Alembertian operator $\Fc_{I}$.
For simplicity to avoid extra complications we assume that these function are
analytic
with real coefficients ${f_{I}}_n$ in Taylor series expansion
$\Fc_{I}=\sum_{n\geq0}{f_{I}}_n\Box^n/M_*^{2n}$. The new mass scale determines
the characteristic scale of the gravity modification. We assume it universal for
all $\Fc_I$ and refer the reader to \cite{BMS} for a detailed discussion of this
new physics parameter. Also apart from the canonical usage of the Riemann tensor
we use 
the Weyl tensor $C^\mu_{\alpha\nu\beta}$ which is coming from the Ricci decmposition
\begin{equation*}
C^{\mu\alpha}_{\phantom{\alpha}\nu\beta}= R^{\mu\alpha}_{\phantom{\alpha}\nu\beta} -\frac 12(\delta^\mu_\nu R^\alpha_{\beta}-\delta^{\mu}_\beta R^{\alpha}_{\nu}+R^\mu_\nu{\delta}^{\alpha}_{\beta}-R^\mu_\beta{\delta}^\alpha_\nu
 )+\frac R{6} (\delta^\mu_\nu {\delta}^{\alpha}_{\beta}-\delta^{\mu}_\beta {\delta}^{\alpha}_{\nu})
\end{equation*}
In this formula we use slightly unusual position of indexes which is useful in performing further computations.
The reason to use the Weyl tensor is because $C^\mu_{\alpha\nu\beta}=0$ on a
conformally flat manifold which is the case for the FLRW metric. We are focused
on the FLRW cosmologies and thus will benefit out of this. Indeed, it means that
the Weyl tensor squared does not show up in the background at all and only
becomes relevant in perturbations. Moreover, even in perturbations the only
non-vanishing contribution is the one where both Weyl tensors are perturbed and
the non-local functions $\Fc_4$ takes its background form.

This action appears in \cite{BGKM}\footnote{It is eq. (27) there. We have
however less terms because possible contractions of the covariant derivatives
with the Ricci tensor can be eliminated thanks to the Bianchi identity or
converted to the higher order in curvature terms using the commutation relations
for the covariant derivatives.} and is the most general covariant non-local
action up to the square in curvature terms and analytic operator
functions $\Fc_I$.\footnote{Also questions
of a ghost-free gravity modification are addressed in \cite{Barv2} using
different action still using
non-local operators constructed out of the d'Alembertian
operator.}
Furthermore we note that working in $4$ dimensions we can assume ${f_2}_0=0$
because using that the Gauss-Bonnet scalar is a total derivative and combining
this with the Ricci decomposition we can have only $R^2$ when no d'Alembertian
operators are in between. In other words among the terms without d'Alembertian
operator  insertions only $R^2$ survives on the FLRW backgrounds. This does not
work for non-constant terms in $\Fc_{I}$ though. 

Equations of motion for action (\ref{model}) can be derived by a straightforward
variation and are as follows
\begin{equation}
\begin{split}
&[M_P^2+2\lambda\Fc_1(\Box)R]G^\mu_\nu={T}^\mu_\nu-\Lambda \delta^\mu_{\nu}
-\frac{\lambda}{2}
 R \Fc_1(\Box) R\delta^\mu_{\nu}+2\lambda(\nabla^\mu\pd_\nu-\delta^\mu_{\nu}
\Box)\Fc_1(\Box) R-\\
-&2\lambda R^\mu_{\beta}\Fc_2(\Box)R^\beta_\nu+\frac\lambda 2\delta^\mu_{\nu}
R^\alpha_\beta\Fc_2(\Box)R^\beta_\alpha+\\
+&
2\lambda\left(\nabla_\rho\nabla_\nu\Fc_2(\Box)R^{\mu\rho}
-\frac12\Box\Fc_2(\Box)R^\mu_\nu-
\frac12\delta^\mu_\nu\nabla_\sigma\nabla_\rho\Fc_2(\Box)R^{\sigma\rho}\right)+\\
+&\lambda{\Kc_1}^\mu_\nu
-\frac{\lambda}{2}\delta^\mu_{\nu}\left({\Kc_1}^\sigma_\sigma+\bar\Kc_{1}
\right)+\lambda{\Kc_2}^\mu_\nu
-\frac{\lambda}{2}\delta^\mu_{\nu}\left({\Kc_2}^\sigma_\sigma+\bar\Kc_{2}
\right)+
2\lambda\Delta^\mu_\nu+2\lambda\Cc^\mu_\nu
\end{split}
\label{EOM}
\end{equation}
where we have defined:
\begin{equation*}
{\Kc_1}^\mu_\nu=\sum_{n=1}
^\infty
{f_1}_n\sum_{l=0}^{n-1}\pd^\mu  R^{(l)}  \pd_\nu  R^{(n-l-1)},~
\bar\Kc_1=\sum_{n=1}
^\infty
{f_1}_n\sum_{l=0}^{n-1} R^{(l)}    R^{(n-l)},
\end{equation*}
\begin{equation*}
{\Kc_2}^\mu_\nu=\sum_{n=1}
^\infty
{f_2}_n\sum_{l=0}^{n-1}\nabla^\mu {R^{(l)}}^\alpha_\beta  \nabla_\nu 
{R^{(n-l-1)}}^\beta_\alpha,~
\bar\Kc_2=\sum_{n=1}
^\infty
{f_2}_n\sum_{l=0}^{n-1} {R^{(l)}}^\alpha_\beta {R^{(n-l)}}^\beta_\alpha,
\end{equation*}
\begin{equation*}
\Delta^\mu_{\nu}=
\sum_{n=1}
^\infty
{f_2}_n\sum_{l=0}^{n-1}\nabla_\beta[ {R^{(l)}}^\beta_\gamma
\nabla^\mu{R^{(n-l-1)}}^\gamma_\nu-\nabla^\mu {R^{(l)}}^\beta_\gamma
{R^{(n-l-1)}}^\gamma_\nu]
\end{equation*}
and $\Cc^\mu_\nu$ is the part coming from the variation of the Weyl tensor
squared piece. As far as we mostly concerned about FLRW cosmologies we simply
neglect this contribution in the background as it is zero. We shall return to
it later during the consideration of perturbations.
Here $G^\mu_\nu$ is the Einstein tensor, $R^{(n)}=\Box^nR$,
${R^{(n)}}^\alpha_\beta=\Box^nR^\alpha_\beta$,  and $T^\mu_\nu$ is the matter
stress tensor if one is present in the system.

One would benefit considering first the trace equation
\begin{equation}
\begin{split}
-&M_P^2R={T}-4\Lambda
-6\lambda\Box\Fc_1(\Box) R-\lambda({\Kc_1} +2\bar\Kc_{1})-\\
-&\lambda\Box\Fc_2(\Box)R- 2\lambda\nabla_\rho\nabla_\mu\Fc_2(\Box)R^{\mu\rho}
-\lambda({\Kc_2} +2\bar\Kc_{2}
)+
2\lambda\Delta+2\lambda\Cc
\end{split}
\label{EOMtrace}
\end{equation}
where quantities without indices denote the trace and significant
simplifications are obvious.

\section{Exact solutions based on the recursion relations}

\subsection{Recursion relations}

In what follows in this Section we use quite intensively ideas and the knowledge
accumulated in \cite{last} (see also \cite{BKM}) where this model without
$\Fc_2$ and $\Fc_4$ pieces was considered and solutions (and their construction)
were discussed. We therefore omit extensive citing of these papers and do refer
the reader to those manuscripts for a detailed analysis of a model which is
$\Fc_2=\Fc_4=0$ in our notations.

Clearly those equations have as solutions Minkowski (if $\Lambda=0$) and de
Sitter (if $\Lambda=\const>0$) backgrounds. The question is however whether we
can find other solutions especially focusing on the possible bouncing
backgrounds.
Equations are definitely highly non-trivial but a decent progress was achieved
in \cite{BMS} by employing the ansatz
\begin{equation}
\Box R=r_1 R+r_2
\label{ansatz}
\end{equation}
in the absence of $\Fc_2$ and $\Fc_4$. This makes the system much simpler
creating the recursive relation for $\Box^n$ and one can try to find exact
solutions.
Known exact solutions (apart from $M_4$ and $dS_4$) include
\begin{equation}
a(t)=a_0\cosh(\sigma t)\text{ found in \cite{BMS}}
\label{sol}
\end{equation}
and
\begin{equation}
a(t)=a_0\exp\left(\frac{\sigma}{2}t^2\right)\text{ found in
\cite{KV_BounceSOl}.}
\label{sol2}
\end{equation}
where $a(t)$ is the scale factor of the FLRW metric
$g_{\mu\nu}=\diag(-1,a(t)^2,a(t)^2,a(t)^2)$.

For the Ricci tensor term we suggest  the following relation as a quite general
ansatz 
\begin{equation}
\Box R^{(n)}_{\mu\nu}=\sum_{m=0}^{n}({r_1}_mR^{(m)}_{\mu\nu}+{r_2}_mg_{\mu\nu}
R^{(m)})+rg_{\mu\nu}
\label{ansatz2pre}
\end{equation}
where ${r_{1,2}}_m$ and $r$ are constants. Surely we require $n$ to be finite.
This would manifest the recursion relation when acting by extra $\Box$-s and
give us a chance to have this as an exact solution to the equations (\ref{EOM}).
Needless to say that smaller $n$ makes the succeeding analysis simpler.

The explanation why finding of such a recursive relation is almost enough is
based on the following arguments. One should start with the trace equation
(\ref{EOMtrace}). If there is no matter or only radiation (which is conformal
and traceless) is present than the trace equation contains only the
gravitational sector of the model.
It can be solved simply by imposing null conditions on coefficients in front of
different powers of the curvature. These are some conditions on the model
parameters but clearly the restrictions are very weak as far as $n$ is finite.
Further we must check one of the component equations (we shall do for $(00)$
component) in (\ref{EOM}) whether it is satisfied or not. But thanks to the
Bianchi identity constraint any discrepancy we can meet would be of the form of
the radiation energy density, i.e. $\sim a(t)^{-4}$. Hence the only question
would be how much radiation must we inject in the system and would it have good
or ghost sign of the energy density.

Of course the proposed above ansatz is of extreme generality and in order to be
more specific we take its following simplified version into the consideration
\begin{equation}
\Box^n{\tilde{G}}^\mu_{\nu}=\alpha_n\Box{\tilde{G}}^\mu_\nu+\beta_n{\tilde
{G}}^\mu_\nu,~n\geq2,\text{~where~}{\tilde{G}}
^\mu_\nu=R^\mu_\nu-\frac14\delta^\mu_\nu R
\label{ansatz2}
\end{equation}
The recursion relation for $n=2$ can be easily solved\footnote{We thank at this
point Dan Thompson for illuminating this issue.} to give
\begin{equation}
\alpha_n=c_+s_+^n+c_-s_-^n,~s_\pm=\frac{\alpha_2\pm\sqrt{\alpha_2^2+4\beta_2}}2,
~
\beta_n=\alpha_{n-1}\beta_2
\label{recursion}
\end{equation}
and $c_\pm$ are to be found examining explicit values of $\alpha_3,~\beta_3$.
However there is a limiting value $n=4$. $n\leq4$ gives a hypothetical chance to
solve the recursion relation while greater $n$ will result in general in
algebraic equations of a degree higher than $4$.

\subsection{Exact analytic bounce}

This choice of the simplified ansatz is motivated by the fact that for solution
(\ref{sol}) we have after some simplifications
\begin{equation}
\Box^2{\tilde{G}}^\mu_{\nu}=14\sigma^2\Box{\tilde{G}}^\mu_\nu-40\sigma^4{\tilde
{G}}^\mu_\nu
\label{ansatz2sol}
\end{equation}

Below we mainly study the particular solution (\ref{sol}) meaning that we have
specific $s_\pm,~c_\pm$. One can, surely, track all the steps considering
(\ref{ansatz2}) in general without specifying particular values for $\alpha_2$
and $\beta_2$. This is interesting in case more solutions satisfying this ansatz
can be found. Going general one would come to a bit more complicated expressions
than we will. We do not do this at the moment for the sake of clarity. Moreover
(\ref{ansatz2}) is already the simplified version of (\ref{ansatz2pre}) which is
worth to analyse if a really general scenario is of interest. 
We thus focus mostly on analysing found solution (\ref{sol}).
 
On solution (\ref{sol}) we find
\begin{equation*}
\alpha_n=\frac1{6\sigma^2}(s_1^n-s_2^n),~\beta_n=-\frac{40\sigma^2}{6}(s_1^{n-1}
-s_2^{n-1}),~s_1=10\sigma^2,~s_2=4\sigma^2.
\end{equation*}
Note that this relations can be used for $n=1$ and $n=0$ as they generate
correct values $\alpha_1=1,~\beta_1=0$ and $\alpha_0=0,~\beta_0=1$ respectively
which are reasonable. Even though ${f_2}_0=0$ as explained in the previous
Section we shall use this property formally to simplify the evaluation of the
sums in the equations of motion.
Apart from this the first ansatz (\ref{ansatz}) is also satisfied with
$r_1=2\sigma^2,~r_2=-24\sigma^4$. We thus have
\begin{equation*}
\Box^n R=r_1^n(R+r_2/r_1),~\Box^n \tilde{G}^\mu_\nu=
\frac{s_1^n}{6\sigma^2}\left(\Box\tilde{G}^\mu_\nu-s_2\tilde{G}^\mu_\nu\right)-
\frac{s_2^n}{6\sigma^2}\left(\Box\tilde{G}^\mu_\nu-s_1\tilde{G}^\mu_\nu\right)
\end{equation*}
These relations yield 
\begin{equation*}
\begin{split}
{\Kc_1}^\mu_\nu=&\Fc^{(1)}_1(r_1)\pd^\mu R\pd_\nu R,\\
\bar\Kc_1=&r_1\Fc^{(1)}_1(r_1)R^2+2r_2\Fc^{(1)}_1(r_1)R
-(\Fc_1(r_1)-{f_1}_0)R\frac{r_2}{r_1}+\\
+&
\frac{r_2^2}{r_1}\Fc^{(1)}_1(r_1)-(\Fc_1(r_1)-{f_1}_0)(r_2/r_1)^2,\\
\end{split}
\end{equation*}
\begin{equation*}
\begin{split}
{{\Kc}_2}^\mu_\nu=&\Fc^{(1)}_2(s_1)\nabla^\mu {S_1}^\alpha_\beta\nabla_\nu
{S_1}_\alpha^\beta+\Fc^{(1)}_2(s_2)\nabla^\mu {S_2}^\alpha_\beta\nabla_\nu
{S_2}_\alpha^\beta-\\
-&\frac{\Fc_2(s_1)-\Fc_2(s_2)}{s_1-s_2}
(\nabla^\mu {S_1}^\alpha_\beta\nabla_\nu {S_2}_\alpha^\beta+\nabla^\mu
{S_2}^\alpha_\beta\nabla_\nu {S_1}_\alpha^\beta)+\frac14\Fc^{(1)}_2(r_1)\pd^\mu
R\pd_\nu R,\\
\bar\Kc_2=&s_1\Fc^{(1)}_2(s_1){S_1}^\alpha_\beta{S_1}_\alpha^\beta+
s_2\Fc^{(1)}_2(s_2){S_2}^\alpha_\beta{S_2}_\alpha^\beta-
\frac{\Fc_2(s_1)-\Fc_2(s_2)}{s_1-s_2}(s_1+s_2){S_1}^\alpha_\beta{S_2}
_\alpha^\beta+\\
+&\frac14\left(r_1\Fc^{(1)}_2(r_1)R^2+2r_2\Fc^{(1)}_2(r_1)R-\Fc_2(r_1)R\frac{r_2
}{r_1}+
\frac{r_2^2}{r_1}\Fc^{(1)}_2(r_1)-\Fc_2(r_1)(r_2/r_1)^2\right),
\end{split}
\end{equation*}
\begin{equation*}
\begin{split}
{\Delta}^\mu_\nu=&\nabla_\beta[ \Fc^{(1)}_2(s_1)({S_1}^\beta_\gamma
\nabla^\mu{S_1}^\gamma_\nu
-\nabla^\mu {S_1}^\beta_\gamma {S_1}^\gamma_\nu)+
\Fc^{(1)}_2(s_2)({S_2}^\beta_\gamma \nabla^\mu{S_2}^\gamma_\nu
-\nabla^\mu {S_2}^\beta_\gamma {S_2}^\gamma_\nu)-\\
&~-\frac{\Fc_2(s_1)-\Fc_2(s_2)}{s_1-s_2}({S_1}^\beta_\gamma
\nabla^\mu{S_2}^\gamma_\nu
-\nabla^\mu {S_1}^\beta_\gamma {S_2}^\gamma_\nu+
{S_2}^\beta_\gamma \nabla^\mu{S_1}^\gamma_\nu
-\nabla^\mu {S_2}^\beta_\gamma {S_1}^\gamma_\nu)
],
\end{split}
\end{equation*}
\begin{equation*}
\begin{split}
\Fc_1(\Box)R=\Fc_1(r_1)R+(\Fc_1(r_1)-f_0)\frac{r_2}{r_1},~
\Fc_2(\Box)R^\mu_\nu=\Fc_2(s_1){S_1}^\mu_\nu-\Fc_2(s_2){S_2}
^\mu_\nu+\frac14\delta^\mu_\nu\Fc_2(\Box)R
\end{split}
\end{equation*}
where we have defined
\begin{equation*}
{S_1}^\mu_\nu=\frac{\Box\tilde{G}^\mu_\nu-s_2\tilde{G}^\mu_\nu}{6\sigma^2},~{S_2
}^\mu_\nu=\frac{\Box\tilde{G}^\mu_\nu-s_1\tilde{G}^\mu_\nu}{6\sigma^2}.
\end{equation*}
The slightly counter-intuitive remixed position of subscripts $1,2$ is correct
here.
$\Fc_{1,2}^{(1)}$ is the derivative with respect to the argument.
In order to cancel in the trace equation structures containing
$\Box\tilde{G}^\mu_\nu$ and $(\Box\tilde{G}^\mu_\nu)^2$ we have to impose the
following conditions
\begin{equation}
\Fc_{2}^{(1)}(s_1)=\Fc_{2}^{(1)}(s_2)=\Fc_{2}(s_1)-\Fc_{2}(s_2)=0
\label{conds1s2}
\end{equation}
and to cancel structures containing $(\pd_\mu R)^2$ and $R^2$
\begin{equation}
\Fc_{1}^{(1)}(r_1)+\frac14\Fc_{2}^{(1)}(r_1)=0.
\label{condr1r2}
\end{equation}
All other equations are also simplified on imposing the above conditions.
In particular we have
\begin{equation*}
\begin{split}
\Fc_2(\Box){R}^\mu_\nu=&\Fc_2(s_1)
(G^\mu_\nu+\frac14\delta^\mu_\nu R)+\frac14\delta^\mu_\nu\Fc_2(r_1)(R+r_2/r_1)
\end{split}
\end{equation*}
or equivalently
\begin{equation*}
\begin{split}
\Fc_2(\Box){\tilde G}^\mu_\nu=&\Fc_2(s_1)
{\tilde G}^\mu_\nu.
\end{split}
\end{equation*}
Careful substitution
in the trace of the Einstein equations (\ref{EOMtrace}) reads
\begin{equation}
\begin{split}
-&M_P^2R={T}-4\Lambda
-6\lambda(\Fc_1(r_1)+\frac14\Fc_2(r_1))(
r_1R+r_2)-\frac\lambda2\Fc_2(s_1)(r_1R+r_2)+\\
+&2\lambda(\Fc_1(r_1)+\frac14\Fc_2(r_1)-{f_1}_0)\frac{r_2}{r_1}(R+r_2/r_1)
\end{split}
\label{EOMtraceansatz}
\end{equation}
and we solve  it (assuming $T=0$) by imposing
\begin{equation}
\begin{split}
\frac{M_P^2}{r_1}-\frac\lambda2\Fc_2(s_1)-6\lambda(\Fc_1(r_1)+\frac14\Fc_2(r_1))
+2\lambda(\Fc_1(r_1)+\frac14\Fc_2(r_1)-{f_1}_0)\frac{r_2}{r_1^2}&=0\\
-\frac{M_P^2}4\frac{r_2}{r_1}&=\Lambda
\end{split}
\label{ansatzsol}
\end{equation}
These relations in terms of $\sigma$ are as follows
\begin{equation}
\begin{split}
\frac{M_P^2}{2\sigma^2}-\frac\lambda2\Fc_2(4\sigma^2)-18\lambda\Fc_1(2\sigma^2)-
\frac92\lambda\Fc_2(2\sigma^2)+12\lambda{f_1}_0&=0\\
3{M_P^2}\sigma^2&=\Lambda
\end{split}
\label{ansatzsolsigma}
\end{equation}
Even though some tuning is required for functions $\Fc_{1,2}$ the solution to
the above conditions is really ambiguous.

Substituting the obtained results in $(00)$ component of (\ref{EOM}) and
performing a number of manipulations we find that only terms proportional to a
constant, $1/\cosh(\sigma t)^2$ and $1/\cosh(\sigma t)^4$ are present and
moreover upon applying the above conditions (\ref{ansatzsolsigma}) one is left
with the following combination
\begin{equation}
\begin{split}
&\rho_r=-\frac{54\lambda\sigma^4}{\cosh(\sigma
t)^4}\left(\Fc_1(r_1)+\frac14\Fc_2(r_1)+\frac1{12}\Fc_2(s_2)\right)
\end{split}
\label{EOMansatz00}
\end{equation}
where we clearly see that we are really left with terms resembling the radiation
(i.e. $\sim a(t)^{-4}$). What is remarkable here that there are two possible
resolutions to have the equations solved completely.
\\
First way is to avoid extra radiation requiring
\begin{equation}
\Fc_1(r_1)+\frac14\Fc_2(r_1)+\frac1{12}\Fc_2(s_2)=0
\label{solnorad}
\end{equation}
This means that extra parameter adjustment is important which however does not
seem to be unnatural. Equations do not trivialize in this case and we can
analyse such a configuration consistently. Note, it is not the case if
$\Fc_2=0$ (see \cite{last} for details).
\\
Second way is to admit some amount of radiation with a weaker requirement
\begin{equation}
\Fc_1(r_1)+\frac14\Fc_2(r_1)+\frac1{12}\Fc_2(s_2)<0
\label{solrad}
\end{equation}
so that this radiation has the positive energy and is not a ghost.

We therefore see that solution (\ref{sol})
equipped with conditions (\ref{conds1s2},\ref{condr1r2}) features a
number of very nice properties: (i) it has non-singular bounce, (ii) it comes at
large time to the de Sitter phase rather than eternal  superinflation, (iii) it
requires just few fine-tunings in the system which still leave us with an
enormous freedom and (iv) it does not have to be supported by extra sources if
we impose an extra condition (\ref{solnorad}) (to avoid this extra tuning we
would assume (\ref{solrad}) to avoid ghosts in the system).

\section{More analytic cosmological solutions?}

In this Section we provide several solutions which unfortunately in most cases
do not
contribute to the question of the non-singular bounce  but are nevertheless
interesting for the future developments in this type of theories.

\subsection{Closely related solutions}

Apart from solution (\ref{sol}) analyzed in the previous Section one can
find after simple analysis that the following are also analytic solutions:
\begin{subequations}
\begin{eqnarray}
a(t)&=&a_0\sinh({\sigma t})\label{sol11}\\
a(t)&=&a_0\cos(\sigma t)\label{sol13}
\end{eqnarray}
\end{subequations}

They can be quite simply obtained using the already known one and
unfortunately none of them contributes to our main subject of a non-singular
bounce. Indeed

\begin{itemize}
\item Solution (\ref{sol11}) is clearly not a bouncing solution and we mention
it here just for the sake of completeness.
\item Solution (\ref{sol13}) is just our already explored solution (\ref{sol})
with an imaginary parameter $\sigma$. It has periodic zeros for the scale
factor. It can be used as a test-bed for discussing possible stages of the
Universe evolution if they obey $\cos(\sigma t)$ behaviour. 
\end{itemize}

Obviously there should be other solutions in the system. At first let us mention
that (\ref{sol2}) does
not generate a recursive relation for the Ricci tensor terms. One could think
also that there is a possibility to solve equation (\ref{EOMtraceansatz}) just
as a differential equation without adjusting the coefficients but this turns out
to be incompatible with the ansatzes. Indeed, equation (\ref{EOMtraceansatz})
was obtained using (\ref{ansatz}) and (\ref{ansatz2}) while a solution to it
does not satisfy these ansatzes identically. Even just finding a solution  other
then presented in the previous Section to the
ansatzes conditions is not an easy task as
it comes to solving at least the third order non-linear differential equation.
As of now it is an open question how to construct more solutions in this model
using some ansatz (see \cite{branko} for various proposal regarding solving
the ansatz condition).

So on one hand we absolutely understand that taming all the non-localities by
virtue of a
recursion relation is the good technical point of the all considered above
solutions. This also must
help in studying perturbations around those particular backgrounds. On the other
hand however this significantly
reduces the possible solutions.
We thus want to go beyond the recursion relations and to find a more general
approach to the problem of solving equations of motion.

\subsection{Model reformulation using ${\tilde{G}}^\mu_\nu$}

The first useful technical step is a passage from the Ricci tensor to the
traceless analog of the Einstein tensor ${\tilde{G}}^\mu_\nu$. We mention in
this regard that appearance of a
combination $\Fc_1+\frac14\Fc_2$ is not spontaneous because we can rewrite the
initial action (\ref{model}) in terms of ${\tilde{G}}^\mu_\nu$ as follows
\begin{equation}
S=\int
d^4x\sqrt{-g}\left(\frac{M_P^2}2R+\frac{\lambda}2\left(R{\tilde
\Fc}_1(\Box)
R+{\tilde{G}}^\mu_\nu\Fc_2(\Box){\tilde{G}}^\nu_\mu+C_{\mu\nu\alpha\beta}
\Fc_4(\Box)C^{\mu\nu\alpha\beta}\right)-\Lambda\right)
\label{tmodel}
\end{equation}
where we have used that ${\tilde{G}}^\mu_\nu$ is traceless and have defined
$\Fc_1(\Box)+
\frac14\Fc_2(\Box)=\tilde{\Fc}_1(\Box)$.
Equations of motion for action (\ref{tmodel}) can be derived by substituting
$R^\mu_\nu={\tilde G}^\mu_\nu+\frac14 \delta^\mu_\nu R$ in (\ref{EOM})
\begin{equation}
\begin{split}
&M_P^2G^\mu_\nu={T}^\mu_\nu-\Lambda \delta^\mu_{\nu}
-2\lambda
 {\tilde G}^\mu_\nu \tilde \Fc_1(\Box)
R+2\lambda(\nabla^\mu\pd_\nu-\delta^\mu_{\nu}
\Box)\tilde \Fc_1(\Box) R-\frac12\lambda R\Fc_2(\Box){\tilde G}^\mu_\nu-\\
-&2\lambda {\tilde G}^\mu_{\beta}\Fc_2(\Box){\tilde G}^\beta_\nu+\frac\lambda
2\delta^\mu_{\nu} {\tilde G}^\alpha_\beta\Fc_2(\Box){\tilde G}^\beta_\alpha+\\
+&
2\lambda\left(\nabla_\rho\nabla_\nu\Fc_2(\Box){\tilde
G}^{\mu\rho}-\frac12\Box\Fc_2(\Box){\tilde G}^\mu_\nu-
\frac12\delta^\mu_\nu\nabla_\sigma\nabla_\rho\Fc_2(\Box){\tilde
G}^{\sigma\rho}\right)+\\
+&\lambda{\Lc_1}^\mu_\nu
-\frac{\lambda}{2}\delta^\mu_{\nu}\left({\Lc_1}^\sigma_\sigma+\bar\Lc_{1}
\right)+\lambda{\Lc_2}^\mu_\nu
-\frac{\lambda}{2}\delta^\mu_{\nu}\left({\Lc_2}^\sigma_\sigma+\bar\Lc_{2}
\right)+
2\lambda\tilde\Delta^\mu_\nu+2\lambda\Cc^\mu_\nu
\end{split}
\label{tEOM}
\end{equation}
where we have defined:
\begin{equation*}
{\Lc_1}^\mu_\nu=\sum_{n=1}
^\infty
{{\tilde f}_1}{}_n\sum_{l=0}^{n-1}\pd^\mu  R^{(l)}  \pd_\nu  R^{(n-l-1)},~
\bar\Lc_1=\sum_{n=1}
^\infty
{\tilde f_1}{}_n\sum_{l=0}^{n-1} R^{(l)}    R^{(n-l)},
\end{equation*}
\begin{equation*}
{\Lc_2}^\mu_\nu=\sum_{n=1}
^\infty
{f_2}_n\sum_{l=0}^{n-1}\nabla^\mu {{\tilde {G}}^{(l)} }{}^\alpha_\beta 
\nabla_\nu  {{\tilde{G}}^{(n-l-1)}}{}^\beta_\alpha
,~\bar\Lc_2=\sum_{n=1}
^\infty
{f_2}_n\sum_{l=0}^{n-1}{\tilde{G}^{(l)}}{}^\alpha_\beta
{\tilde{G}^{(n-l)}}{}^\beta_\alpha,
\end{equation*}
\begin{equation*}
\tilde\Delta^\mu_{\nu}=
\sum_{n=1}
^\infty
{f_2}_n\sum_{l=0}^{n-1}\nabla_\beta[ {\tilde{G}^{(l)}}{}^\beta_\gamma
\nabla^\mu{\tilde{G}^{(n-l-1)}}{}^\gamma_\nu-\nabla^\mu
{\tilde{G}^{(l)}}{}^\beta_\gamma {\tilde{G}^{(n-l-1)}}{}^\gamma_\nu]
\end{equation*}
and ${\tilde f_1}{}_n$ are coefficients of the Taylor expansion of function
$\tilde \Fc_1$.
The Weyl tensor related part may have an impact now since we are going to
consider perturbations. One can find the relevant part of it is
\begin{equation*}
\Cc^\mu_\nu=\left(R_{\alpha\beta}
+2\nabla_\alpha\nabla_\beta\right)
\Fc_4(\Box)C_{\nu}^{\phantom{\nu}\alpha\beta\mu}.
\end{equation*}
Saying relevant we mean only the piece which is obtained by the variation of
one of the Weyl tensor factors in the action. Then we are left with only
one Weyl tensor as it is obvious from the latter formula and further
perturbation of this remaining Wel tensor factor may produce a non-zero
contribution to the perturbation equations.

The trace equation becomes
\begin{equation}
\begin{split}
-&M_P^2R={T}-4\Lambda
-6\lambda\Box\tilde\Fc_1(\Box) R-\lambda({\Lc_1} +2\bar\Lc_{1})-\\
-& 2\lambda\nabla_\rho\nabla_\mu\Fc_2(\Box){\tilde G}^{\mu\rho}
-\lambda({\Lc_2} +2\bar\Lc_{2}
)+
2\lambda\tilde\Delta
\end{split}
\label{tEOMtrace}
\end{equation}
and the Weyl tensor related term $\Cc$ turns out to be zero thanks to the full
tracelessness of the Weyl tensor.

This form of action and equations of motion also turns out to be beneficial for
studying perturbations as we will see in the next Section.

\subsection{Avoiding recursion relations}

Surely we are still sticking to the FLRW type of the metric but keep up to some
extent the scale factor general. This means that we cannot say anything specific
about functions $R^{(n)}$ which appear after the box operator acts on the scalar
curvature $n$ times. Fortunately, we now can have a tiny progress with the
second rank tensor ${\tilde G}^\mu_\nu$. It is traceless and action of the box
operator does not break this property because box commutes with the metric. It
means that
${\tilde{G}^{(n)}}{}^\mu_\mu=0$ for any $n$ and thus we can introduce
\begin{equation}
\Box^n {\tilde G}^\mu_\nu=b_n(t){\zeta}^\mu_\nu
\text{ where }
\zeta^\mu_\nu=\diag(3,-1,-1,-1)
\label{solsqrtRmunu21}
\end{equation}
where one can compute that
\begin{equation}
b_{n+1}(t)=\left(\Box+8H^2\right)b_n(t)\text{ and }
b_0(t)=\dot H/2.
\label{solsqrtRmunu22}
\end{equation}
Here as usual $H=\dot a/a$ is the Hubble function  and dot is the derivative
with respect to the cosmic
time $t$.

This allows us to simplify all the tensor structures in the trace equation
(\ref{tEOMtrace}). As was explained in the previous Section this is the
cornerstone equation since a solution to it is almost automatically a solution
to
all the Einstein equations modulo perhaps some radiation (assuming no other
matter is present in the system).
Careful substitution gives
\begin{equation}
\begin{split}
\Lc_2&=12\sum_{n=1}^\infty {f_2}_n\sum_{l=0}^{n-1}\left(\pd^\mu b_l(t)\pd_\mu
b_{n-l-1}(t)-8H^2b_l(t)b_{n-l-1}(t)\right),\\
{\bar \Lc}_2&=12\sum_{n=1}^\infty {f_2}_n\sum_{l=0}^{n-1}b_l(t)b_{n-l}(t),\\
{\tilde\Delta}&=-48\sum_{n=1}^\infty {f_2}_n\sum_{l=0}^{n-1}\left(
H\pd_t(b_l(t)b_{n-l-1}(t))+(3H^2+\dot H)b_l(t)b_{n-l-1}(t))
\right),\\
\nabla_\rho\nabla_\mu\Fc_2(\Box){\tilde G}^{\mu\rho}&=
3\Box b(t)-12 H(\dot b(t)+Hb(t))\text{ where
}b(t)=\sum_{n=1}^\infty({f_2}_nb_n).
\end{split}
\label{EOMtracesqrtbna}
\end{equation}
It is manifest now that we have reduced the problem of solving the trace
equation to the problem of many scalar functions and moreover all of them are
connected by means of the covariant box operator. To have the complete picture
we note down
\begin{equation}
R=12H^2+6\dot H.
\label{scalarR}
\end{equation}

Possible recursion relations may be of course substituted here. Technically
presence of a recursion relation for $R^{(n)}$ or $b_n$ or both means that the
corresponding infinite series recurses after a certain $n$. The following
configurations may appear than in this model:

\begin{enumerate}

\item
If both $R^{(n)}$ and $b_n$ obey some recursion conditions for a given $a(t)$
then this in turn means that there is a finite set of linearly independent
functions and in order to satisfy the trace Einstein equation we must equate to
zero coefficients in front of each such function. For a finite $n$ this produces
finite number of conditions on ${f_{1,2}}_k$. This is exactly what is
implemented in the previous Section for the solution (\ref{sol}).

\item
If only, say, $R^{(n)}$ obeys a recursion relation while $b_n$ not then this
means that in general infinite fine tuning is required. Indeed, absence of
recursion for $b_n$ is equivalent to the fact that infinitely many linear
independent functions will be generated all accompanied by ${f_2}_n$
coefficients. Here we see two ways to satisfy the trace Einstein equation.
Either all the coefficients ${f_2}_n$ must be trivial eliminating the
corresponding action modification at all or we must fine tune all of them such
that the infinite series of functions sums up to a constant. The latter
possibility may be accomplished in the spirit of the so called ``addition
theorem'' for the associate Legendre functions when some certain infinite series
sums up to the unit (see for instance formulae (8.794) and (8.814)
in \cite{GradsteinRyzhik}). For this to happen it is vital to have infinite
number of
non-zero coefficients ${f_2}_n$. This is reasonable in our set-up but on the
other hand all of them must be fine tuned. This is not definitely unacceptable,
however. When, for example, we speak about a
function like $\Fc_2=e^{\beta\Box}$ we say it has one free parameter $\beta$
rather than infinitely many. The point here is that these operator functions
come normally from the consideration of the corresponding string models as
mentioned in the Introduction and their nature is therefore originates from some
other principles and not from our cosmological model. Thus in the case some
solution needs an infinite number of coefficients to be fixed we should only
worry whether the resulting operator function is acceptable from the point of
view of the string model it comes from. Example with such properties is solution
(\ref{sol2}).

\item
If neither $R^{(n)}$ nor $b_n$ appear to obey a recursion all the coefficients
${f_{1,2}}_n$ are in the play. Again they can be either trivialized or fine
tuned. Here, however another type of fine tuning may arise when instead of
fixing all the coefficients we do fix only some relations in between of $f_1$-s
and $f_2$-s. This is still an infinite fine tuning from the point of view of
individual coefficients but from the point of view of the operator functions
$\Fc$ this may be just one simple relation. To be more precise, provided
cancellation of all terms happens if ${f_1}_n/{f_2}_n=\const$ which does not
depend on
$n$ then it is equivalent to have $\tilde{\Fc}_1\sim\Fc_2$.

\end{enumerate}

\subsection{Other solutions}

\textbf{Solutions of the first type.}

Solution belonging to the first type when both the scalar curvature and
the
Ricci tensor obey recursion relations, apart from (\ref{sol}), is
\begin{equation}
a(t)=a_0t
  \label{sol3}
\end{equation}
One can find that on this solution 
\begin{equation*}
R=6/t^2,~\Box R=0,~b_0=-\frac1{2t^2},~b_1=-\frac4{t^2},~b_2=0
\end{equation*}
and it is easy to show that equations of motion can be satisfied. It is not a
bouncing solution but may be interesting in analyzing the regime when the scale
factor grows linearly.

$~$

\noindent\textbf{Configurations of the second type.}

Known configurations of the second type are as follows
\begin{itemize}
  \item 
\begin{equation}
a(t)=a_0\sqrt{t}
  \label{sol4}
\end{equation}
For this scale factor we have 
\begin{equation*}
R=0,~b_0=-\frac1{4t^2},~b_1=\frac1{4t^2},~b_2=-\frac3{t^6},~\dots
\end{equation*}
and sequence of $b$-s does not stop.
  \item
Equations become a bit more transparent if we pass to $a(t)$ as the
variable. We can
do this in each region where we have one to one correspondence $t\leftrightarrow
a(t)$. We therefore can rewrite equations in the new variable and time
derivatives transform to $\pd_t=aH(a)\pd_a$. The most important for us
d`Alembertian operator
$\Box$ becomes
\begin{equation*}
\Box=-a^2H^2\left(\pd_a^2+\left(\frac4a+\frac{H'}{H}\right)\pd_a\right)
\end{equation*}
where prime is the derivative with respect to $a$.
Another important operator is
\begin{equation*}
\Box+8H^2=-a^2H^2\left(\pd_a^2+\left(\frac4a+\frac{H'}{H}\right)\pd_a-\frac8{a^2
}\right)
\end{equation*}
since it creates functions $b_n$.

To have a connection with the previously obtained solution note that
\begin{equation*}
H=\sigma\sqrt{1-\frac{a_0^2}{a^2}}
\end{equation*}
gives $a(t)=a_0\cosh(\sigma t)$.

Than one can guess some interesting configurations and one of them is
\begin{equation*}
H=\sigma\sqrt{1-\frac{a_0^4}{a^4}}
\end{equation*}
giving
\begin{equation}
a(t)=a_0\sqrt{\cosh(\sigma t)}
\label{sol5}
\end{equation}
We have for this scale factor
\begin{equation*}
  R=3\sigma^2
\end{equation*}
while $b$-s form an infinite sequence of polynomials of $\cosh(\sigma t)$.

One interesting observations is that the scale factor given by (\ref{sol2}) is
exactly the leading term in the series expansion
\begin{equation*}
H=\sigma\sqrt{1-\frac{a_0^p}{a^p}}
\end{equation*}
in the limit $p\to 0$. Also note that for scale factor (\ref{sol2}) we have
\begin{equation*}
\begin{split}
 R&=12\sigma^2t^2+6\sigma,~\Box R=-72\sigma^3t^2-24\sigma,~
  \dots\\
 b_0&=\sigma/2,~b_1=4\sigma^2t^2,~\dots
\end{split}
\end{equation*}
where we see explicitly that $R$ obeys a recursion relation while $b$ not.

In all these cases one need an infinite fine tuning as explained in the previous
Subsection in order to perform summations in the equations of motion. Even
though
such a fine tuning may be not favorable the reformulation itself of equations of
motion in terms of $a$
may be interesting and we are looking forward to go further in this direction.
\end{itemize}
Note also that all the above configurations of the second type are
solutions with
$\Fc_2=0$.

$~$

\noindent\textbf{Configurations of the third type.}

The third type of configurations is extremely generic and the only known scale
factor which makes us able to track equation up to some reasonable extent is
\begin{equation}
a=a_0t^p
  \label{sol6}
\end{equation}
for a generic $p$. For this scale factor we have
\begin{equation*}
\begin{split}
 R^{(l)}&=\frac{r_l}{t^{2l+2}},~
 b_l=\frac{\beta_l}{t^{2l+2}}
\end{split}
\end{equation*}
where constant coefficients may be expressed through $p$ explicitly. Note that
for odd $p$ the series for $R^{l}$ stops at a certain $l$ and the soluions
falls back in the previous subclass when one recursion relation exists. Now it
is a matter of a straightforward computation to see that one can satisfy the
Einstein equations by imposing algebraic relations on coefficients ${f_1}_n$
and ${f_2}_n$. This is because internal summations over $l$ in expressions for
$\Lc$-s do not involve the time variable $t$.

$~$

Note also that all the above solutions are still solutions with an arbitrary
$\Fc_4$. Even though it is not important for the background one would change
the perturbative picture playing with the $\Fc_4$ parameter.

\section{A road-map to perturbations}

Needless to say that even writing down the perturbation equations may be
an unfeasible task in the above formulation of the model. Below in this
Section we outline the general approach and emphasize what shall
be developed shortly in subsequent papers since the full detailed analysis is
expected
to be rather cumbersome and deserves a separate paper.

\subsection{Bianchi identity}

We know that the Einstein equations are constrained with the Bianchi identity
which says $\nabla_\mu G^\mu_\nu\equiv 0$. In our case we have more than just
Einstein-Hilbert Lagrangian but all the additional ingredients we have are
covariant terms. This guarantees that the Bianchi identity holds trivially
without imposing any extra condition. On the other hand this implies thanks to
arbitrariness of coefficients ${f_I}_n$ that each separate term does
covariantly conserve. Indeed, each ${f_I}_n$ is a coefficient in front of
some covariant structure in the Einstein equations, say $\tau^\mu_\nu$. Assuming
that only one of $f$-s coefficients is non-zero we come to a conclusion that the
corresponding structure must covariantly conserve due to
Bianchi identities, i.e. $\nabla_\mu\tau^\mu_\nu\equiv 0$. In other words it
resembles a conserving perfect fluid stress-energy tensor. The same argument is
applicable to all the $f$-s coefficients as well as their arbitrary
combinations.

The above arguments imply that thanks to Bianchi identity the parts with
different $\Fc_I$
covariantly conserve separately. To make use of this we define
\begin{equation*}
  \begin{split}
{T_0}^\mu_\nu&={T}^\mu_\nu,\\
{T_1}^\mu_\nu&=-2\lambda
 {\tilde G}^\mu_\nu \tilde \Fc_1(\Box)
R+2\lambda(\nabla^\mu\pd_\nu-\delta^\mu_{\nu}
\Box)\tilde \Fc_1(\Box) R+\lambda{\Lc_1}^\mu_\nu
-\frac{\lambda}{2}\delta^\mu_{\nu}\left({\Lc_1}^\sigma_\sigma+\bar\Lc_{1}
\right),\\
{T_2}^\mu_\nu&=
-\frac12\lambda R\Fc_2(\Box){\tilde G}^\mu_\nu
-2\lambda {\tilde G}^\mu_{\beta}\Fc_2(\Box){\tilde G}^\beta_\nu+\frac\lambda
2\delta^\mu_{\nu} {\tilde G}^\alpha_\beta\Fc_2(\Box){\tilde G}^\beta_\alpha+\\
&+2\lambda\left(\nabla_\rho\nabla_\nu\Fc_2(\Box){\tilde
G}^{\mu\rho}-\frac12\Box\Fc_2(\Box){\tilde G}^\mu_\nu-
\frac12\delta^\mu_\nu\nabla_\sigma\nabla_\rho\Fc_2(\Box){\tilde
G}^{\sigma\rho}\right)+\\
&+\lambda{\Lc_2}^\mu_\nu
-\frac{\lambda}{2}\delta^\mu_{\nu}\left({\Lc_2}^\sigma_\sigma+\bar\Lc_{2}
\right)+
2\lambda\tilde\Delta^\mu_\nu,\\
{T_4}^\mu_\nu&=2\lambda\left(R_{\alpha\beta}
+2\nabla_\alpha\nabla_\beta\right)
\Fc_4(\Box)C_{\nu}^{\phantom{\nu}\alpha\beta\mu}.
\end{split}
\end{equation*}
Now Einstein equations can be written in an extremely concise form
\begin{equation}
\begin{split}
M_P^2G^\mu_\nu&=\sum_I{T_I}^\mu_\nu-\Lambda \delta^\mu_{\nu}\\
\end{split}
\label{tEOMT}
\end{equation}
and moreover we have
\begin{equation*}
\begin{split}
\nabla_\mu{T_I}^\mu_\nu=0\text{ for any }I.
\end{split}
\end{equation*}

One recognizes here the system of minimally coupled perfect fluids minimally
coupled to gravity. The perturbation technique is  known in general (see
\cite{hwangnoh,KV} for instance) but it is not obvious it can be
applicable ``as is'' to our model as we will see shortly.

\subsection{Subtleties of the non-localities}

Having brought Einstein equations in our model to an already studied structure
does not mean we are able easily, if at all, solve the perturbation equations.

It looks promising that we could clearly separate the whole equations into 
minimally coupled terms. But we have to stress here that each term is not a
canonical perfect fluid, i.e. it cannot be written just in terms of energy
density $\rho$, pressure density $p$ and four-fluid velocity $u_\mu$ as
discussed in
\cite{Mukhanov_rev}.\footnote{In other words we cannot just have
${T_I}^\mu_\nu=(\rho+p)u^\mu u_\nu-p\delta^\mu_\nu$.} This results in presence
of anisotropic stresses as well as entropic perturbations for our fluids at the
level of perturbations. This in turn does not allow us to use the already known
techniques straightforwardly since the system of perturbation equations (even
non-local ones)
will not be closed.

On the other hand all $T_I$ are not independent external stress energy tensors
but rather some structures built on the metric. We therefore must be able in
principle to have the corresponding perturbative quantities in terms of the
metric perturbations.

Also one must not go straight with $T_4$ term since it represents a
contribution which is absent in the background. It means, that energy and
pressure for $T_4$ are both zero. Thus it is not obvious how one can define the
equation of state parameter $w=p/\rho$ and the speed of sound ${c_s}^2=\dot
p/\dot\rho$.

It is, however, the simplest term for performing actual calculations at the
perturbative level. This is because it is zero on the background and completely
traceless. The first point means that
\begin{equation*}
\begin{split}
\delta{T_4}^\mu_\nu&=2\lambda\left(R_{\alpha\beta}
+2\nabla_\alpha\nabla_\beta\right)
\Fc_4(\Box)\delta C_{\nu}^{\phantom{\nu}\alpha\beta\mu}
\end{split}
\end{equation*}
The second point means that for isotropic perturbations
\begin{equation*}
\begin{split}
\delta
C_{\nu}^{\phantom{\nu}\alpha\beta\mu}=c(\eta)e^{i\vec
k\vec x}P_{\nu}^{\phantom{\nu} \alpha\beta\mu}
\end{split}
\end{equation*}
where $\eta$ is the conformal time, $k$ is the comoving wave-vector and
$P_{\nu}^{\phantom{\nu}
\alpha\beta\mu}$ is a constant tensor which keeps
all the symmetry properties of the Weyl tensor.
We thus can use the same idea as in the
previous Section for the traceless tensor ${\tilde{G}}{}^\mu_\nu$ and compute
an action of some differential operator $\Dc$ on $\delta
C_{\nu}^{\phantom{\nu}\alpha\beta\mu}$ as
\begin{equation*}
\begin{split}
\Dc\delta
C_{\nu}^{\phantom{\nu}\alpha\beta\mu}=(\tilde{\Dc}
c(\eta))e^{i\vec k\vec x}P_{\nu}^{\phantom{\nu}
\alpha\beta\mu}
\end{split}
\end{equation*}
with a simple algebraic relation between $\Dc$ and $\tilde\Dc$.

We can push these arguments further and draw some conclusions when $R$ or
${\tilde{G}}{}^\mu_\nu$ are perturbed in $T_1$ and $T_2$. We then are able to
compute implicitly action of the non-local operators on those perturbed
quantities. However, the real problem comes would one perturb the non-local
operators themselves and compute the action of this perturbed operators on
unperturbed background quantities. This is where presence of recursion
relations is crucial. This was the key point why could we end up with final
results in \cite{last}. It is an open question how one would go in a general
case when there are no recursion relations. We hope to address this issue in
the forthcoming research \cite{next}.

We nevertheless can come up with a conclusion that the discussed above solution
(\ref{sol}) is an example of a stable bounce configuration. Even though this
particular solution creates recursion relations and in principle one would much
easier write down system of perturbation equations we can claim the bounce
phase is stable without tedious calculations. This stability is guaranteed
because solution has analytic dependence on coefficients $f$. Stability of
this solution with $\Fc_2=\Fc_4=0$ was explicitly demonstrated in \cite{last}.
Therefore turning on $\Fc_2$ and/or $\Fc_4$ must keep the configuration stable
up to some range of coefficients ${f_{2,4}}_n$. Of course, this argument is not
enough to find the allowed domain for new coefficients and this is the primary
goal for the forthcoming paper \cite{next}.

\subsection{de Sitter limit}

The de Sitter limit is perhaps the most simple configuration for analysis of
perturbations (apart from Minkowskian space-time, of course). The de Sitter
Universe is described by the scale factor
\begin{equation*}
  a=a_0e^{H t}
\end{equation*}
and the corresponding Hubble function is just the constant $H$. This is where we
see why the traceless tensor ${\tilde{G}}^\mu_\nu$ is convenient since it is
identically zero in such a metric. Moreover the scalar curvature is a constant
$R=12H^2$. One can easily check this is a solution to our equations. But the
most important that the part of equations which may be relevant for
perturbations is much shorter
since we can drop terms quadratic in ${\tilde G}^\mu_\nu$ as well as some
others. Then the relevant equations become
\begin{equation}
\begin{split}
&M_P^2\delta G^\mu_\nu=\delta {T}^\mu_\nu
-2\lambda
 \delta {\tilde G}^\mu_\nu \tilde {f_1}_0
R+2\lambda(\nabla^\mu\pd_\nu-\delta^\mu_{\nu}
\Box)\tilde \Fc_1(\Box) \delta R-\frac12\lambda R\Fc_2(\Box)\delta {\tilde
G}^\mu_\nu-\\
+&2\lambda\left(\nabla_\rho\nabla_\nu\Fc_2(\Box)\delta {\tilde
G}^{\mu\rho}-\frac12\Box\Fc_2(\Box)\delta {\tilde G}^\mu_\nu-
\frac12\delta^\mu_\nu\nabla_\sigma\nabla_\rho\Fc_2(\Box)\delta {\tilde
G}^{\sigma\rho}\right)-\\
-&\frac{\lambda}{2}\delta^\mu_{\nu}R\left(\tilde \Fc_1(\Box) -\tilde
{f_1}_0
\right)\delta R+4\lambda\nabla_\alpha\nabla_\beta
\Fc_4(\Box)\delta C_{\nu}^{\phantom{\nu}\alpha\beta\mu}
\end{split}
\label{tEOMds}
\end{equation}
with the trace
\begin{equation}
\begin{split}
-&M_P^2\delta R=\delta {T}
-6\lambda\Box\tilde\Fc_1(\Box) \delta R-2\lambda R\left(\tilde \Fc_1(\Box)
-\tilde
{f_1}_0
\right)\delta R- 2\lambda\nabla_\rho\nabla_\mu\Fc_2(\Box)\delta {\tilde
G}^{\mu\rho}.
\end{split}
\label{tEOMtraceds}
\end{equation}

These can be analyzed up to the end explicitly for some configurations.
To highlight the track towards solutions to the perturbation equations for
scalar perturbations we note that $\delta R$ has an explicit expression in
terms of Bardeen potentials $\Phi$ and $\Psi$, the two gauge invariant scalar
degrees of freedom \cite{Bardeen}. Then one can show that $i\neq j$ component of
equation (\ref{tEOMds}) is a non-local equation containing $\delta R$ and 
$\Phi-\Psi$, where $i,j$ are the spatial indexes. Assuming further that
the matter is the radiation without anisotropic stresses and computing
explicitly 
$\nabla_\rho\nabla_\mu\Fc_2(\Box)\delta {\tilde
G}^{\mu\rho}$ which is feasible thanks to the tracelessness of the ${\tilde
G}^{\mu\rho}$ tensor we must end up with two non-local equations on $\Phi$ and
$\Psi$.

Even
though we can develop on this right away we are going to stop at this stage 
because there are many other serious and
unrevealed questions regarding the de Sitter background in this type of models
such as the ghost-free condition. These
unexplored matters deserve a separate study which will include the detailed
analysis of perturbations as well  \cite{next}.


\section{Summary and Outlook}

We have considered in this paper the most general extension of GR based on
inclusion of stringy motivated non-local operators and keeping the quadratic
in curvature terms.

The primary goal was to find out an exact bouncing solution and it is for
the first time as this task is accomplished up to the best of our knowledge. It
is intriguing that the bouncing configuration is the cosine hyperbolic
which is still a solution provided operator functions $\Fc_2$ and $\Fc_4$ which
control the presence of extra terms are trivial. This indicates a presence of
some symmetry one would have to find and so far is one of the open
questions.
A good point of our bouncing solution (\ref{sol}) is that it may be a solution
in our model without extra matter which was not the case without $\Fc_2$ and
$\Fc_4$. Also this solution requires just a couple of conditions which are very
general and are easy to satisfy.

Moreover we have presented several other solutions which are not all
bouncing solutions but anyway widen our understanding of the model. Practically
we divided possible solutions into three groups based on the fact whether they
create or not
recursion relations when the scalar curvature and the Ricci tensor are
being acted by
powers of the d'Alembertian operator. Among all the found configurations
the one which is (\ref{sol5}) looks very interesting for
the further study since it creates a bouncing Universe with the constant scalar
curvature. It is a solution with $\Fc_2=0$ or for a very special $\Fc_2$ when
all the coefficients are strictly fixed.

One more interesting point of this model is that the last term with the Weyl
tensor squared in the Lagrangian does not contribute to the background at all.
It, however, shows up in perturbations and it is interesting to see whether
presence of this term can be efficiently used to tackle the problem of solving
the perturbation equations.

Also we were able to bring the model formulation to such a form that was
already analyzed at the perturbative level for other configurations. The
perturbations can be analyzed
in full at least numerically in the de Sitter limit and the clear way towards
this was outlined. As it was mentioned in the main part of the paper, however,
we postpone
the more detailed and technical study of this question since it is more natural
to join it with other unrevealed problems one can put for this model in the de
Sitter space. Namely, ghost-free condition  and subsequent quantization are
the issues to be fully
studied as well and results should appear shortly
in~\cite{next}.

The full analysis of perturbations is expected to be very hard in general. At
the moment it is not even obvious one will be able to close the system of
perturbation equations for scalar perturbations. Tensor perturbations may
become simpler but still one would face the difficulty that effective
anisotropic stresses present in the system. This is tough and open question at
the moment.

Furthermore one question which we kept aside is the anisotropic
perturbations during the contraction phase since this question is under
investigation in the parallel project~\cite{nextKV}. The generic expectation is
that such perturbations must grow during the contraction phase and the main
problem is to formulate the domains of parameters of the model allowing the
bounce to happen smoothly. Approaches to study anisotropic
perturbations in non-local as well as stringy-inspired models can
be found in \cite{oobBI,BI} and references therein.


\acknowledgments

A.K. is supported by an ``FWO-Vlaanderen''
postdoctoral fellowship and also
supported in part by Belgian Federal Science
Policy Office through the Interuniversity Attraction Pole P7/37, the
``FWO-Vlaanderen'' through the project G.0114.10N and the RFBR grant
11-01-00894.


\end{document}